\documentclass[english]{article}
\usepackage[T1]{fontenc}
\usepackage[latin9]{inputenc}
\usepackage{amsmath}
\usepackage{amsthm}
\usepackage{amssymb}

\makeatletter
\numberwithin{equation}{section}
\numberwithin{figure}{section}
\newcommand{\lyxaddress}[1]{
\par {\raggedright #1
\vspace{1.4em}
\noindent\par}
}

\makeatother

\usepackage{babel}
\begin{document}

\title{\textbf{Einstein and Rastall Theories of Gravitation in Comparison}}
\maketitle

\subsection*{F. Darabi$^{1,2}$, H. Moradpour$^{1}$, I. Licata$^{3,4}$, }

\subsection*{Y. Heydarzade$^{1,2}$ and C. Corda$^{1}$}

\medskip{}

\medskip{}

\medskip{}

\lyxaddress{\textbf{$^{1}$}Research Institute for Astronomy and Astrophysics
of Maragha (RIAAM), Maragha 55134-441, Iran}

\lyxaddress{$^{2}$Department of Physics, Azarbaijan Shahid Madani University,
Tabriz, Iran}

\lyxaddress{$^{3}$ISEM, Inst. For Scientific Methodology, PA, Italy \& $^{4}$School
of Advanced International Studies on Applied Theoretical and Non Linear
Methodologies in Physics, Bari (Italy) }

\lyxaddress{Correspondence: $cordac.galilei@gmail.com$}
\begin{abstract}
We profit by a recent paper of Visser claiming that Rastall gravity
is equivalent to Einstein gravity to compare the two gravitational
theories in a general way. Our conclusions are different from Visser's
ones. We indeed argue that these two theories are not equivalent.
In fact, Rastall theory of gravity is an \textquotedbl{}open\textquotedbl{}
theory when compared to Einstein general theory of relativity. Thus,
it is ready to accept the challenges of observational cosmology and
quantum gravity.
\end{abstract}
\begin{quotation}
\textbf{Keywords: Rastall gravity; Einstein gravity; non-minimal coupling;
conservation laws.}
\end{quotation}
\begin{quote}
\textbf{PACS numbers: 04.20.-q; 04.50.-h; 04.50.Kd.}
\end{quote}
The framework of extended theories of gravity {[}1 - 4, 44{]} is today
considered an useful and popular approach to attempt to solve the
important problems of the standard model of cosmology like Dark Energy
\cite{key-5,key-6} and Dark Matter \cite{key-7,key-8}. A key point
is that all of the potential alternatives to Einstein's general theory
of relativity (GTR) must be viable. This implies that alternative
theories must be in agreement with the Einstein's equivalence principle,
which has today a strong, unchallengeable empirical evidence \cite{key-2}.
The fundamental consequence is that the alternatives to Einstein gravity
must be metric theories \cite{key-2}. Another important point is
that such alternatives must pass the solar system tests. Hence, deviations
from the standard GTR must be weak \cite{key-1,key-2}. 

Within the framework of extended gravity, one finds the theory proposed
by P. Rastall in 1972 \cite{key-9}, which recently obtained a renewed
interest in the literature {[}10 - 13{]}. This interest is due to
some good behavior of the Rastall theory. In fact, on one hand it
seems in good agreement with observational data on the Universe age
and on the Hubble parameter \cite{key-14}. On the other hand it may,
in principle, provide an alternative description for the matter dominated
era with respect to the GTR \cite{key-15}. Observational data from
the helium nucleosynthesis seem also in agreement with Rastall gravity
\cite{key-16}. Based on these observational evidences, there have
been some recently studies of the various cosmic eras in the framework
of Rastall gravity {[}17 - 21{]}. Other interesting issues are that
Rastall gravity seems to do not suffer from the entropy and age problems
of standard cosmology \cite{key-22} and is consistent with the gravitational
lensing phenomena \cite{key-23,key-24}. Further interesting works
on Rastall gravity are given by {[}25 - 29{]} and references within.

Differently from Einstein gravity, Rastall gravity considers a non-divergence-free
energy-momentum {[}9 - 29{]}. A different, but similar theory is the
so called \emph{Curvature-matter theory of gravity }{[}30 - 34{]}.
In this theory the matter and geometry are coupled to each other in
a non-minimal way\emph{ }{[}30 - 34{]}. Then, the standard energy-momentum
conservation law does not work in this case too {[}30 - 34{]}. 

In a recent paper \cite{key-35}, Visser claimed that Rastall gravity
is equivalent to Einstein gravity. In this letter, we argue that these
two theories are not equivalent instead. Let us see the key points
of our interpretation.

Rastall did not define a new energy-momentum tensor (EMT) in his original
paper \cite{key-9}. He only assumed that the ordinary conservation
law is not always valid in curved space-time and proposed a model
for it. In fact, he proposed a new relation between EMT and geometry
which is supported by an unknown mutual interaction between them.
We think that the ordinary definition of EMT is valid in Rastall's
hypothesis and there is an unknown non-minimal coupling between geometry
and matter fields in Rastall gravity which should be found out in
order to write its Lagrangian. Hence, the Lagrangian for the Rastall
theory should be written as 
\begin{equation}
L_{E}+L_{M}+L_{i},\label{eq: Lagrangiana totale Rastall}
\end{equation}
where $L_{E}$ is the standard Lagrangian of the GTR, which arises
from the Ricci scalar $R,$ $L_{M}$ is the ordinary Lagrangian of
the matter fields, and $L_{i}$ is a term which includes a non-minimal
coupling between geometry and matter fields. This is exactly the term
which leads to the Rastall term ($Rg_{ab}$). This is an unknown term
which seems to need more investigations and, in general, can be a
function of $L_{E},$ $L_{M},$ $R,$ $T,$ $G_{ab}$ and $T_{ab}.$
In summary, it seems incorrect to say that Rastall's definition of
EMT is not true, as Visser has claimed in the third section of his
paper \cite{key-35}. We think indeed that the ordinary definition
of EMT is valid in the Rastall theory, unlike the claims of Visser
in \cite{key-35}. In fact, we stress that violation of energy conservation
has been invoked also in frameworks which are different from the Rastall
gravity approach, for example in a recent attempt to achieve Dark
Energy \cite{key-36}, while a study of classical fields in the background
of a fluctuating space-time volume is considered as being a particular
case of Rastall gravity \cite{key-37}.

We disagree with the reasoning introduced by Visser in the third section
of \cite{key-35}. In fact, Visser starts from a modified theory of
gravity (Rastall theory) and uses the gravitational field equations
to rewrite the field equations in a way in which the geometrical part
is the same as that of the GTR and the matter part differs from that.
Then, comparing the rewritten equations with the GTR equations, Visser
gets a relation between the ordinary EMT and the EMT of Rastall gravity
(Eqs. (12 - 15) in \cite{key-35}). In other words, Visser implicitly
assumes a priori that the Rastall equation, i.e. Eq. (2) in \cite{key-35},
is equivalent to the standard Einstein's equation, i.e. Eq. (9) in
\cite{key-35}. As a consequence, all Visser's derivations in the
third Section of \cite{key-35} are simple rearrangements, but, in
our opinion, such rearrangements have no physical meaning being due
to Visser's a priori assumption that Eqs. (2) and (9) in \cite{key-35}
are equivalent. They are not equivalent instead. Let us see this point
in more detail. The key point is that in \cite{key-35} Visser has
put a new tensor, labelled $\left[T_{R}\right]_{ab}$ in the right
hand side of his Eq. (2), whereas if one rearranges the terms in Eq.
(1) in Rastall's original paper \cite{key-9} , one sees that it has
the form 
\begin{equation}
G_{ab}=\kappa T_{ab}^{(m)}-\kappa\lambda Rg_{ab},\label{eq: Rastall original}
\end{equation}
where $T_{ab}^{(m)}\neq\left[T_{R}\right]_{ab}$ is the ordinary EMT.
Now, if one uses the normalization of Visser $(\kappa\lambda=\lambda/4)$,
one finds 
\begin{equation}
G_{ab}+\frac{\lambda}{4}Rg_{ab}=\kappa T_{ab}^{(m)}.\label{eq: Rastall original 2}
\end{equation}
Comparing this equation (originally obtained by Rastall in \cite{key-9}
without Visser's normalization $\kappa\lambda=\lambda/4$) with Eq.
(2) of Visser's paper \cite{key-35} which is 
\begin{equation}
G_{ab}+\frac{\lambda}{4}Rg_{ab}=\kappa\left[T_{R}\right]_{ab},\label{eq: Visser mistake}
\end{equation}
one realizes that Visser has erroneously put $\left[T_{R}\right]_{ab}$
instead of $T_{ab}^{(m)}$ in the right hand side of Eq. (\ref{eq: Rastall original 2}).
If he would have written the correct $T_{ab}^{(m)}$, then he would
have obtained the correct eqs. (8) and (9) in \cite{key-35}, in which
$\left[T_{R}\right]_{ab}$ is replaced by $T_{ab}^{(m)}$. Thus, he
would have obtained $G_{ab}=\kappa\left[T_{R}\right]_{ab}$ instead
of $G_{ab}=\kappa T_{ab}^{(m)}$. In fact, the original version of
Rastall gravity \cite{key-9} leads to the modified equation 
\begin{equation}
G_{ab}=\kappa T^{(m)}R{}_{ab},\label{eq: Rastall Ricci}
\end{equation}
(where $\ensuremath{\left[T_{R}\right]_{ab}=(T_{ab}^{(m)}+\frac{1}{4}\frac{\lambda}{1-\lambda}T^{(m)}g_{ab})}$)
and NOT 
\begin{equation}
G_{ab}=kT_{ab}^{(m)},\label{eq: Einstein standard}
\end{equation}
 (where $\ensuremath{T_{ab}^{(m)}=(\left[T_{R}\right]_{ab}+\frac{1}{4}\frac{\lambda}{1-\lambda}\left[T_{R}\right]g_{ab})}$)
as Visser has obtained. Due to the Bianchi identity one finds 
\begin{equation}
\ensuremath{\nabla^{b}G_{ab}=0},\label{eq: Conservation 1}
\end{equation}
and, in turn, 
\begin{equation}
\ensuremath{\nabla^{b}\left[T_{R}\right]_{ab}=0},\label{eq: Conservation 2}
\end{equation}
which is NOT $\ensuremath{\nabla^{b}T_{ab}^{(m)}=0.}$ Now, Eq. (\ref{eq: Conservation 2})
is equivalent to 
\begin{equation}
\ensuremath{\nabla^{b}T_{ab}^{(m)}=-\frac{1}{4}\frac{\lambda}{1-\lambda}\nabla^{b}T^{(m)}g_{ab}},\label{eq: non Conservation}
\end{equation}
or 
\begin{equation}
\ensuremath{\nabla^{b}T_{ab}^{(m)}=\frac{1}{4}{\lambda}\nabla^{b}Rg_{ab}},\label{eq: non Conservation 2}
\end{equation}
which accounts for the non-conservation of Einstein's EMT $T_{ab}^{(m)}$
according to the original idea of Rastall in \cite{key-9}. Hence,
as the source of $\ensuremath{\nabla^{b}T_{ab}^{(m)}}$ is the 4-dimensional
geometry itself (namely $\ensuremath{\nabla^{b}Rg_{ab}}$), the consequence
is that Rastall gravity is drastically different from Einstein gravity
in 4-dimensions.

Some criticisms on Rastall gravity (and on other non-conservative
gravitational theories) have also been raised in the early paper \cite{key-43}.
In fact, the discussion in \cite{key-43} is almost the same as the
one in \cite{key-35} and a bit more detailed. One verbatim reads
in \cite{key-43}: \emph{``The question which remains is how to construct
the stress-energy tensor out of the matter fields of the real world.
Since $S_{\mu\nu}$ is always exactly conserved, and the definition
of $S_{\mu\nu}$ in terms of $T_{\mu\nu}$ (3), is independent of
space-time curvature, this remaining question is really a question
about the conservation of stress-energy in special relativity, and
not an alternative theory of gravity at all}.'' The quantities\emph{
$S_{\mu\nu}$ }and \emph{$T_{\mu\nu}$ }in \cite{key-43} are the
same quantities that we labelled $\left[T_{R}\right]_{ab}$ and $T_{ab}^{(m)}$
in this letter, respectively. Exactly like in \cite{key-35},\emph{
$S_{\mu\nu}$} is defined as being ``a new stress-energy tensor''
by the authors of \cite{key-43}. There is a key point in the sentence
in \cite{key-43} ``\emph{the definition of $S_{\mu\nu}$ in terms
of $T_{\mu\nu}$ (3), is independent of space-time curvature}'' which
have led the authors of \cite{key-43} (and Visser in \cite{key-35})
to suppose that there is essentially no non-minimal coupling of matter
with geometry. The point is that, there is certainly a non-minimal
coupling present in the coupling $\lambda/1-4\lambda$ in Eq. (3)
of \cite{key-43}. This coupling comes exactly from the trace of the
Rastall field equation. The apparent absence of the Ricci scalar in
Eq. (3) of \cite{key-43} does not mean that it is independent of
the space-time curvature. In fact, the definition of $S_{\mu\nu}$
in terms of $T_{\mu\nu}$ (3) in Eq. (3) of \cite{key-43} IS DEPENDENT
on the space-time curvature indirectly through the trace equation.
Without such dependence and non-minimal coupling between the Ricci
scalar $R$ and the trace of the EMT, the $\lambda/1-4\lambda$ in
Eq. (3) of \cite{key-43} would not exist and so\emph{ $S_{\mu\nu}$}
would not satisfy the conservation equation. Such a non-minimal coupling
is crucial for the covariant conservation of\emph{ $S_{\mu\nu}$ }. 

Visser also argues in \cite{key-35} that since the ordinary EMT differs
from the obtained EMT, the Rastall's definition of EMT is incorrect.
We think that this argument should be not true. If it is true, then
one can generalize his recipe to all of the modified gravity theories
and concludes that all of them are incorrect and are special rearrangements
of the GTR, which is, of course, not true. Let us consider, as a simple
example, $f(R)$ gravity \cite{key-3,key-4}. In that case, the field
equations are \cite{key-4}
\begin{equation}
G_{ab}=\frac{1}{f'(R)}\{\frac{1}{2}g_{ab}[f(R)-f'(R)R]+f'(R)_{;a;b}-g_{ab}\square f'(R)\}+\kappa T_{ab}^{(m)},\label{eq: einstein 2}
\end{equation}
where $\kappa$ is the gravitational coupling constant. Defining the
\emph{curvature fluid} EMT as \cite{key-4} 
\begin{equation}
T_{ab}^{(c)}\equiv\frac{1}{\kappa f'(R)}\{\frac{1}{2}g_{ab}[f(R)-f'(R)R]+f'(R)_{;a;b}-g_{ab}\square f'(R)\},\label{eq: curvatura}
\end{equation}
one immediately rewrites Eq. (\ref{eq: einstein 2}) as 
\begin{equation}
G_{ab}=\kappa T_{ab}^{(c)}+\kappa T_{ab}^{(m)}.\label{eq: einstein 2 compatta}
\end{equation}
Now, if one introduces 
\begin{equation}
T_{ab}^{(total)}\equiv(T_{ab}^{(m)}+T_{ab}^{(c)}),\label{eq: tensore totale}
\end{equation}
the gravitational field equations in $f(R)$ gravity obtain the Einsteinian
form 
\begin{equation}
G_{ab}=\kappa T_{ab}^{(total)}.\label{eq: Einstein modified}
\end{equation}
Eqs. (\ref{eq: Einstein modified}) are formally equal to the traditional
field equations of the GTR, which are 
\begin{equation}
G_{ab}=\kappa T_{ab}^{(m)},\label{eq: Einstein traditional}
\end{equation}
but, of course, this does not mean that the GTR is equivalent to $f(R)$
gravity. In fact, such an equivalence exists only in the particular
case $f(R)=R,$ as it is well known.

It is also important to stress that two different modified theories
may have different vacuum solutions since there may be some different
degrees of freedom associated to each of these theories. Reversely,
if two theories have same vacuum solutions, one can not deduce any
result for the equivalence of those theories. This is because the
vacuum solution is just a specific solution which one can not read
all the degrees of freedom of the underlying theory from that. For
example, in \cite{key-38} it has been shown that there are vacuum
solutions on a 4D brane which are completely different than the vacuum
solutions in the GTR. The same happens for the brane solutions with
non-vanishing bulk Weyl tensor.

We find remarkable differences between Einstein and Rastall gravities
in cosmological solutions, as it has been recently shown in \cite{key-39}.
In addition, we emphasize that, from the cosmological point of view,
Rastall hypothesis (mutual interaction between geometry and matter
fields) cannot describe Dark Energy and one need to consider a Dark
Energy-like source, just the same as Einstein GTR \cite{key-20,key-39}.
In addition, one may describe the current era, by generalizing the
Rastall hypothesis and without using a Dark Energy-like source \cite{key-40}. 

Finally, non-equivalence between Einstein and Rastall theories of
gravity was early stressed in an old but interesting paper of Smalley
\cite{key-41}.

For the sake of completeness, we take the chance to recall some further
interesting issue and some weakness of Rastall gravity. From the point
of view of Mach principle, Rastall gravity seems more ``Machian''
than Einstein gravity \cite{key-42}. Till now, physicists interacting
with Rastall gravity have at most focused on its cosmological features.
Thus, one can consider as being a weakness of the Rastall framework
the difficulty to relate the accelerating expansion of the Universe
to the Rastall interaction. On the other hand, this is also a well
known problem of the GTR. Another weakness is that the Rastall Lagrangian
is not completely known, despite a first approach on this issue has
been recently developed in \cite{key-37}. In fact, we think that
Rastall theory of gravity needs more studies to become familiar with
all its aspects.

\subsection*{Conclusion remarks}

In this letter, we profited by the recent paper of Visser \cite{key-35},
where it is claimed that Rastall gravity is equivalent to Einstein
gravity, to compare the two gravitational theories in a general way.
The conclusions of the current work are different from the ones in
\cite{key-35}. We have indeed raised various issues showing that
these two theories are not equivalent. As a consequence, we can consider
Rastall gravity as being an \textquotedbl{}open\textquotedbl{} theory
compared to Einstein gravity. Hence, Rastall theory of gravity seems
to be an extended theory of gravity ready to accept the challenges
of observational cosmology and quantum gravity.

\subsection*{Acknowledgements}

This letter has been supported financially by the Research Institute
for Astronomy and Astrophysics of Maragha (RIAAM), Project Number
1/5440-15.

\end{document}